\documentclass[aps,amssymb,prl,twocolumn,showpacs]{revtex4}

\usepackage{epsfig}
\usepackage{array}

\def\0{\varnothing}

\begin{document}
\title{Hysteresis in one-dimensional reaction-diffusion systems}
\author{A.~R\'akos, M.~Paessens, G. M. Sch\"{u}tz}
\affiliation{Institut f\"{u}r Festk\"{o}rperforschung, Forschungszentrum
J\"{u}lich, 52425 J\"{u}lich, Germany.}
\date{\today}

\begin{abstract}
We introduce a simple nonequilibrium model for a driven diffusive system
with nonconservative reaction kinetics which exhibits ergodicity breaking
and hysteresis in one dimension. These phenomena can be understood through a 
description of the dominant stochastic
many-body dynamics in terms of an equilibrium single-particle problem,
viz. the random motion of a shock in an
effective potential. This picture also leads to the exact phase diagram of the
system and suggests a new generic mechanism for ``freezing by heating''.
\end{abstract}

\maketitle

The closely related questions of phase coexistence, ergodicity breaking and 
hysteresis in noisy one-dimensional systems with short range interactions and
finite local state space (such as in spin systems or vertex models) are
intriguing. In thermal equilibrium these phenomena cannot occur as there
is no local mechanism that could limit the growth of islands of a minority 
phase inside a majority phase. Far from equilibrium one has found phase 
separation and spontaneous symmetry breaking in driven diffusive systems 
provided that either a bulk conservation law, viz. particle number conservation
\cite{Evan95,Evan98a,Arnd98b,Barl98,Kafr02}, or vanishing local transition 
rates \cite{Helb99,Evan02} constrain the local dynamics. As already noted in 
Ref. \cite{Evan95} the only known exception to this rule, the error-correcting 
model by Gacs \cite{Gacs86}, is rather complicated and still
not widely understood, see also \cite{Gray01}.

Recently it has been demonstrated that phase coexistence occurs in a
one-dimensional driven diffusive system even in the presence of Langmuir
kinetics $A \rightleftharpoons 0$ which break the bulk conservation law
\cite{Parm03}. This mechanism is inspired by the process of motor proteins
moving along actin filaments. Earlier this model was introduced as a toy
model reproducing stylized facts in limit order markets \cite{Will02}.
The formation of a localized shock in this system which
separates a domain of low particle density from a domain of high density
has been studied subsequently \cite{Popk03,Evan03}. However, the two different
domains do not represent two possible {\it global} steady states. The
process is
ergodic even in the thermodynamic limit and no hysteresis is possible.

It is the purpose of this letter to present a simple nonequilibrium system
with local non-conservative dynamics and finite local state space which
exhibits ergodicity breaking and hysteresis in the thermodynamic limit,
in the usual sense that in finite volume the sojourn time in two
metastable steady states increases exponentially with system size. To be
specific we investigate the totally asymmetric exclusion process (TASEP)
augmented by nonconservative reaction kinetics. The TASEP is a stochastic
model of diffusing particles on a one-dimensional lattice with a hopping
bias in one direction \cite{Schu01}.
Each site from 1 to $L$ is either empty or occupied by one particle.
In the bulk particles ('$A$') hop stochastically from
site $i$ to $i+1$ with unit rate, provided that the target site is empty.
The boundaries act as particle reservoirs with densities $\rho_{-}$ on the left
resp. $\rho_{+}$ on the right: On site $1$ particles are created with rate
$\rho_{-}$, provided the site is empty, which corresponds to a particle hopping
from the left reservoir onto the first site. Particles on site $L$ are
annihilated with rate $1-\rho_{+}$, corresponding to a particle hopping from
the last site into the right reservoir.

In our model particles also undergo the following reaction process: On a vacant
site enclosed by two particles a particle may be attached with rate $\omega_a$,
and a particle enclosed by two other particles may be detached with rate
$\omega_d$. This process can be symbolically written as $A\0 A
\rightleftharpoons AAA$ and may be interpreted as activated Langmuir kinetics.
Without the TASEP dynamics the stationary density of this process is either
$K=\omega_a/(\omega_a+\omega_d)$ or zero, with no correlations
\cite{Rako03b}. As in previous work we consider the
physically interesting case when $L\to\infty$ and these rates are proportional
to $1/L$ \cite{Parm03,Will02,Popk03,Evan03}. Hence we define renormalized rates
\begin{equation}
\omega_a=\Omega_a/L, \qquad \omega_d=\Omega_d/L
\end{equation}
where $\Omega_a$ and $\Omega_d$ are kept constant while $L\to\infty$.
For other choices of the
attachment/detachment (AD) rates the dynamics is  either governed by the TASEP
($\omega_{a,d}<{\cal O}(1/L)$)
or by the AD process ($\omega_{a,d}>{\cal O}(1/L)$).

\begin{figure}
\centerline{\epsfig{file=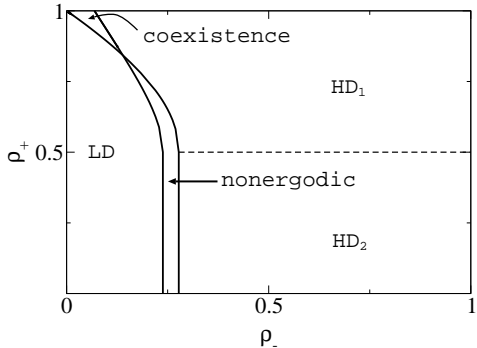,width=7truecm}}
\caption{Phase diagram for $\Omega_a=0.7$ and $\Omega_d=0.1$ with
two high density phases (HD1, HD2),
a low density phase (LD), a coexistence phase and the nonergodic
phase.
}\label{fig1}
\end{figure}

We find a stationary
phase diagram of the model with five distinct phases (Fig.~\ref{fig1}). The
stationary density profile $\rho_i$ is not constant as a function of lattice
site $i$. Yet some of the phases are comparable to those of the usual TASEP
with open boundaries \cite{Schu93b,Derr93b}: in the high
density phases (HD1/2) one finds $\rho_i>1/2$ while in the low density phase
(LD) $\rho_i<1/2$. In HD1 the bulk density profile is
dependent on $\rho_{+}$, while it is independent of both boundaries in HD2 as
in the maximal current phase of the TASEP. On the other hand two additional
phases exist: (i) A coexistence phase which is characterized by a stable shock
position, i.e., a jump in the density profile which is localized at a certain
position in the bulk of the system. The shock connects a low density domain to
its left with a high density domain to its right
as known from related models studied previously \cite{Parm03,Popk03}.
Notice that in the
usual TASEP there is a coexistence line in the phase diagram with a
{\it nonlocalized} shock. In a different parameter
regime we find a novel phase with an unstable shock position in the
bulk.
In this phase both the LD and HD states are stable (if $L \to \infty$)
which implies that ergodicity is
broken in the thermodynamic limit.
Although for finite systems a transition between the two states is
possible, the mean life time of each steady state
is exponentially large in the system size $L$
(see below). We note that this is not a spontaneous symmetry
breaking since
there is no symmetry relating the two metastable states. This phase has no
analog in the TASEP with open boundaries.

\begin{figure}[t]
\centerline{\epsfig{file=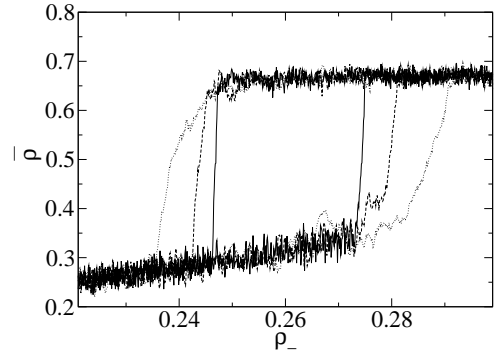,width=5.6truecm,angle=270}}
\caption{Hysteresis plot for $L=2000$, $\Omega_a=0.7$, $\Omega_d=0.1$,
  $\rho_+=0.45$. $\rho_-$ was changed by $10^{-4}$ in every 5000 (solid
  line), 1500 (dashed line) and 500 (dotted line) MC steps. The
  hysteresis loop gets wider as the speed of changing $\rho_-$ is increased.}
\label{fig6}
\end{figure}

Hysteresis in this nonequilibrium setting was observed by measuring
the space-averaged density $\bar\rho$ along the curve of constant $\rho_+=0.45$
while
changing $\rho_-$ in such a way that the system
starting from the LD phase passed through the nonergodic phase and ended up in
the HD2 phase. Then the process of changing $\rho_-$ was reversed.
A relevant parameter in hysteresis phenomena is the
speed of sweeping: in
our simulations $\rho_-$ was changed by $10^{-4}$ in every $k$
MC steps ($k=500,1500,5000$). A time average was not taken, $\bar\rho$ was
measured in
every $k$ steps. On Fig.~\ref{fig6} one can see the resulting hysteresis
loops. We found that
the hysteresis loop inflates with increasing speed which is reminiscent of
hysteresis in usual magnetic systems.

To rationalize these observations we first consider
the hydrodynamic limit on the Euler scale, i.e., we take
$L\to\infty$ while the
lattice constant is scaled by $a=1/L$ and the time by
$t=t_{\text{lattice}}/L$. Thus the spatial
coordinate $x=i/L$ becomes continuous. Following the line of arguments of
Ref.~\cite{Popk03}
the hydrodynamic equation for the density takes the form
\begin{equation}
\frac{\partial}{\partial t}\rho(x,t) +\frac{\partial}{\partial x}j(\rho)=
S(\rho),
\end{equation}
with the exact current $j(\rho)=\rho(1-\rho)$ of the TASEP and the cubic
source term
\begin{equation}
S(\rho)=\Omega_a \rho^2 (1-\rho) - \Omega_d \rho^3.
\end{equation}
resulting from the activated Langmuir kinetics.
For the stationary state $\partial_t\rho(x,t)=0$ holds and
using $\partial_x j={\partial j}/{\partial \rho} \cdot
{\partial\rho}/{\partial x}$
we obtain
\begin{equation}
v_c(\rho)\frac{\partial\rho(x)}{\partial x}=S(\rho),
\label{eq:flowfield}
\end{equation}
with the collective velocity $v_c=\partial j/\partial\rho$.
This nonlinear differential equation can be integrated analytically
and yields the flow field
\begin{equation}
x(\rho)=-\frac{1}{\Omega_a\rho}+\frac{\Omega_a-\Omega_d}{\Omega_a^2}
\ln\left|\frac{1}{K}-\frac{1}{\rho}\right|+c
\label{flowfield}
\end{equation}
with an integration constant $c$.

As the differential equation is of first order and the boundary condition
fixes the density at two positions, following a line of the flow field does
not represent a solution of the boundary problem in general. In the original
lattice model this inconsistency is resolved by the appearance of a
shock and/or boundary layers as described in \cite{Popk03}. Apart from the
discontinuities the stationary density profile follows the flow field of
eq.~(\ref{eq:flowfield}).

In order to understand quantitatively the selection of the stationary shock
position (which determines the phase diagram) and also to explain the
phenomenon of hysteresis
from a microscopic viewpoint we describe the dominant dynamical mode of the
particle system in terms of the random motion of the shock. To this end
we generalize the approach of \cite{Kolo98}
and introduce space-dependent hopping rates
\begin{eqnarray}
w_{x\to x+a} =\frac{j_R(x)}{\rho_R(x)-\rho_L(x)}\nonumber \\
w_{x+a\to x} =\frac{j_L(x)}{\rho_R(x)-\rho_L(x)}.
\label{eq:rates}
\end{eqnarray}
for jumps of the shock over a lattice constant $a$.
Here the indices $L$ and $R$ denote the solutions (lines of the
flow field (\ref{flowfield})) on the left resp. right of the shock.
Similar hopping rates are used in \cite{Evan03}.
The space-dependent hopping rates furnish us with the picture of a random
walker in an effective energy landscape $E(x)$ inside a finite box.
The energy landscape is generated by the interplay of the particle current
with the reaction kinetics. In this way we relate the original nonequilibrium
many-particle system to an equilibrium single-particle model.
Let $p(x)$ be the equilibrium probability of the shock being at position
$x$. Then due to detailed balance
\begin{equation}
\frac{w_{x\to x+a}}{w_{x+a\to x}} = \frac{p(x+a)}{p(x)} = \exp(-E(x+a)+E(x)).
\label{eq:energy}
\end{equation}
which defines the energy landscape.

\begin{figure}[t]
\centerline{\epsfig{file=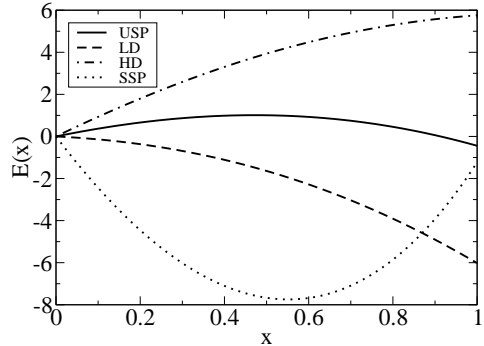,width=5.6truecm,angle=270}}
\caption{Examples for the energy landscape in four phases. Note that
  in the HD and LD phases $E(x)$ can be either convex or concave}
\label{fig5}
\end{figure}

The potential $E(x)$ is monotonically increasing (decreasing) function
for the HD (LD) phase (Fig.~\ref{fig5}). This implies that although there are
fluctuations the shock is always driven to the left (right) boundary.
In the coexistence phase there is a global minimum in the bulk resulting in
a stable shock position (Fig.~\ref{fig5}) at a macroscopic distance from
the boundaries. Here the dynamics can be well
approximated by a random walker in a harmonic potential which gives a
Gaussian distribution for the shock position. Hence the width of the
shock distribution is proportional to $\sqrt{L}$ \cite{Rako03b}
which was also found in \cite{Parm03,Evan03} for the TASEP with
Langmuir kinetics.

The nonergodic phase is characterized by a global energy {\em maximum} in the
bulk (Fig.~\ref{fig5}), leading to an unstable bulk fixed point of the shock.
The two minima at the left and right boundary correspond
to the two stable stationary states. Starting
with an initial condition close to one of the minima, the random walker will
drift most likely into
this local minimum and stay in its vicinity for a time of the order of
the mean first passage time
$\bar{\tau}$ before it traverses to the other minimum. This leads to
hysteresis.
Using a formula for the mean first passage time derived by Murthy and Kehr
\cite{MK}
one expects that $\bar{\tau}$ grows exponentially with the system size
$L$. Moreover, one expects the transition from one
minimum to the other to
be a random Poisson process with an average waiting time $\bar{\tau}$.

\begin{figure}[t]
\centerline{\epsfig{file=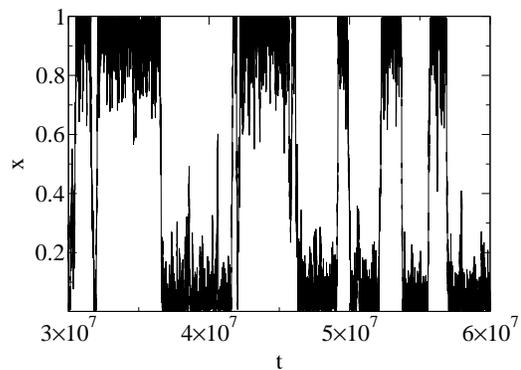,width=5.6truecm,angle=270}}
\caption{Snapshot of the time evolution of the scaled position of the second class particle for $L=1000, \rho_{-}=0.2705,
\rho_{+}=0.63, \Omega_a=0.5, \Omega_d=0.1$. A position of the second
class particle near the
left boundary ($x\approx 0$) corresponds to the high density state,
while a position near the
right boundary ($x\approx 1$) corresponds to the low density state.
}\label{fig3}
\end{figure}

This simple one-particle picture is well borne out by MC simulations.
For judiciously chosen parameters it is possible to perform simulations
up to times much larger than $\bar{\tau}$.
Using multispin coding \cite{Bark} for the MC algorithm rather good statistics
become available
for the waiting time $\tau$ (the time the system spends in one of the
stationary states before switching to the other). For tracing the
position of the shock we use the second class particle technique \cite{Ferr91}.
We measured the position of the second class particle as a function of time:
a typical realization is shown in Fig.~\ref{fig3}.

\begin{figure}[t]
\centerline{\epsfig{file=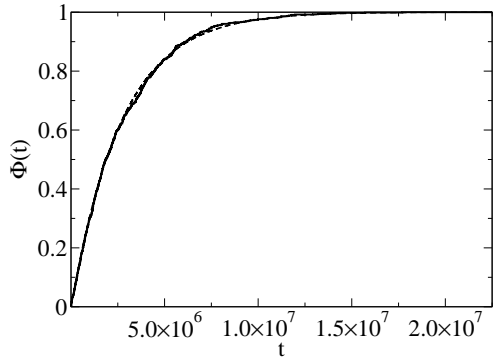,width=5.6truecm,angle=270}}
\caption{Numerically determined cumulative distribution of the transition times
from the upper state to the lower (solid line)
compared to the exponential distribution (dashed line) with
parameters as in \ref{fig3}. Similar results are found for the transition in
the other direction, but with a different $\bar{\tau}$ \cite{Rako03b}.
}\label{fig4}
\end{figure}

As shown in Fig.~\ref{fig4} the numerically determined
cumulative distribution function
$\Phi(t)=P(\tau<t)$ of the waiting time $\tau$
is hardly distinguishable from the expected exponential distribution
\begin{equation}
\Phi(t)=1-\exp(-t/\bar{\tau}).
\end{equation}

With this picture of a moving shock in mind and using the expression
(\ref{flowfield}) it is also possible to derive the exact phase transition
lines defining the phase diagram presented above. Adapting the arguments of
\cite{Popk03} the analysis of the stability of boundary layers
yields a high density phase for $\rho_{-}>1/2$. Because of a boundary layer
the bulk solution of the density profile is independent of $\rho_{+}$ if
$\rho_{+}<1/2$. Thus in this region the phase diagram is only ruled by
$\rho_{-}$. For $\rho_- \leq 1/2$ the two lines in the phase diagram bounding
the coexistence phase and
nonergodic phase resp. are determined by the stationary shock position.
Crossing the line separating LD/coexistence phase and nonergodic/HD phase from
left to right results in a change of the sign of
$\partial_x E(1)$ from $-\to +$. Crossing the other line separating 
coexistence/HD phase and 
LD/nonergodic phase from left to right results in a change of the sign of 
$\partial_x E(0)$
from $-\to +$.
The sign of the slope of the energy profile, i.e., the stability of the shock
position can be analysed
by considering the average shock velocity
\begin{equation}
v_s=\frac{j_R(x)-j_L(x)}{\rho_R(x)-\rho_L(x)}.
\label{eq:vs}
\end{equation}
A shock position at the boundary is stable when it is driven toward the
boundary, i.e., $v_s(0)<0$ at the left, $v_s(1)>0$ at the right boundary. Thus
the lines separating the phases are calculated by comparing the values of
$\rho_L(x)$ and $\rho_R(x)$ at the positions $x={0,1}$.

To conclude we have demonstrated the existence of hysteresis and broken
ergodicity (in the thermodynamic limit) in a driven diffusive system
without bulk conservation law. We stress that the two different stationary
distributions are not ordered states in which the activated Langmuir
reaction kinetics would be dynamically suppressed. Surprisingly, adding noise
which is on average spatially homogeneous (the nonconservative reaction
process) to a conservative spatially homogeneous
nonequilibrium system with a nonvanishing particle current leads to
a {\it space-dependent} effective potential which determines the stationary
position of the shock. In the absence of this noise, i.e., in the
usual TASEP, the shock performs an unbiased random walk and hence is
unlocalized, whereas suitably chosen reaction kinetics may create a variety
of effective potentials which localize the shock. An increase in noise strength
is usually associated with heating up a system whereas localization
reduces the amount of disorder, corresponding to cooling.
Thus we have identified a novel generic mechanism for the phenomenon
of freezing by heating. The description of the nonequilibrium many-body
dynamics in terms of a collective single-particle mode moving under
equilibrium conditions yields the exact stationary phase diagram as well
as the numerically verified flipping process between the metastable
states of the finite system. Details of the flipping dynamics will be
presented elsewhere \cite{Rako03b}.

\begin{acknowledgments}
A. R\'akos thanks the Deutsche Forschungsgemeinschaft (DFG)
for financial support.
\end{acknowledgments}

\end{document}